\documentclass[preprint]{vgtc}               % preprint
\usepackage{mathptmx}
\usepackage{graphicx}
\usepackage{times}

\usepackage{balance}  % to better equalize the last page
\usepackage[T1]{fontenc}
\usepackage{txfonts}
\usepackage{mathptmx}
\usepackage[pdftex]{hyperref}
\usepackage{color}
\usepackage{booktabs}
\usepackage{textcomp}

\onlineid{0}

\vgtccategory{Research}

\vgtcinsertpkg

\title{Effects of Hand Representations for Typing in Virtual Reality}

\author{Jens Grubert\thanks{e-mail: jg@jensgrubert.de}\\ %
        \scriptsize  Coburg University of Applied Sciences and Arts
\and Lukas Witzani\thanks{e-mail: lukas.witzani@uni-passau.de}\\ %
     \scriptsize University of Passau 
\and Eyal Ofek\thanks{e-mail: eyalofek@microsoft.com}\\ %
     \scriptsize Microsoft Research
\and Michel Pahud\thanks{e-mail: mpahud@microsoft.com}\\ %
     \scriptsize Microsoft Research
\and Matthias Kranz\thanks{e-mail: matthias.kranz@uni-passau.de}\\ %
     \scriptsize University of Passau
\and Per Ola Kristensson\thanks{e-mail: pok21@cam.ac.uk}\\ %
     \scriptsize University of Cambridge}

\teaser{
	\centering
	\includegraphics[width=2.1\columnwidth]{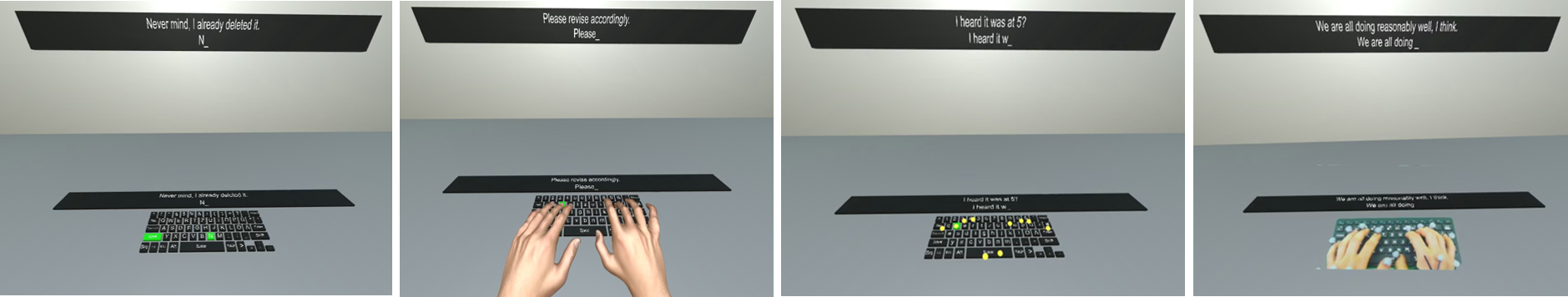}
	\caption{Views on the conditions studied in the experiment on effects of hand representations for typing in VR. From left to right:  \textsc{NoHand}, \textsc{IKHand}, \textsc{Fingertip} and \textsc{VideoHand}. }
	\label{fig:conditionshandstudy}
}
\abstract{
Alphanumeric text entry is a challenge for Virtual Reality (VR) applications. VR enables new capabilities, impossible in the real world, such as an unobstructed view of the keyboard, without occlusion by the user's physical hands. Several hand representations have been proposed for typing in VR on standard physical keyboards. However, to date, these hand representations have not been compared regarding their performance and effects on presence for VR text entry. Our work addresses this gap by comparing existing hand representations with minimalistic fingertip visualization. We study the effects of four hand representations (no hand representation, inverse kinematic model, fingertip visualization using spheres and video inlay) on typing in VR using a standard physical keyboard with 24 participants. We found that the fingertip visualization and video inlay both resulted in statistically significant lower text entry error rates compared to no hand or inverse kinematic model representations. We found no statistical differences in text entry speed.

} % end of abstract

\CCScatlist{ 
  \CCScat{H.5.2:}{ User Interfaces - Input devices and strategies.}{}{}
}

\begin{document}


%\firstsection{Introduction}

\maketitle

\section{Introduction}

%\todo[inline]{TODO LIST IS NOW AT THE END OF THE DOCUMENT}

\begin{figure}[h]
	\centering
	\includegraphics[width=0.8\columnwidth]{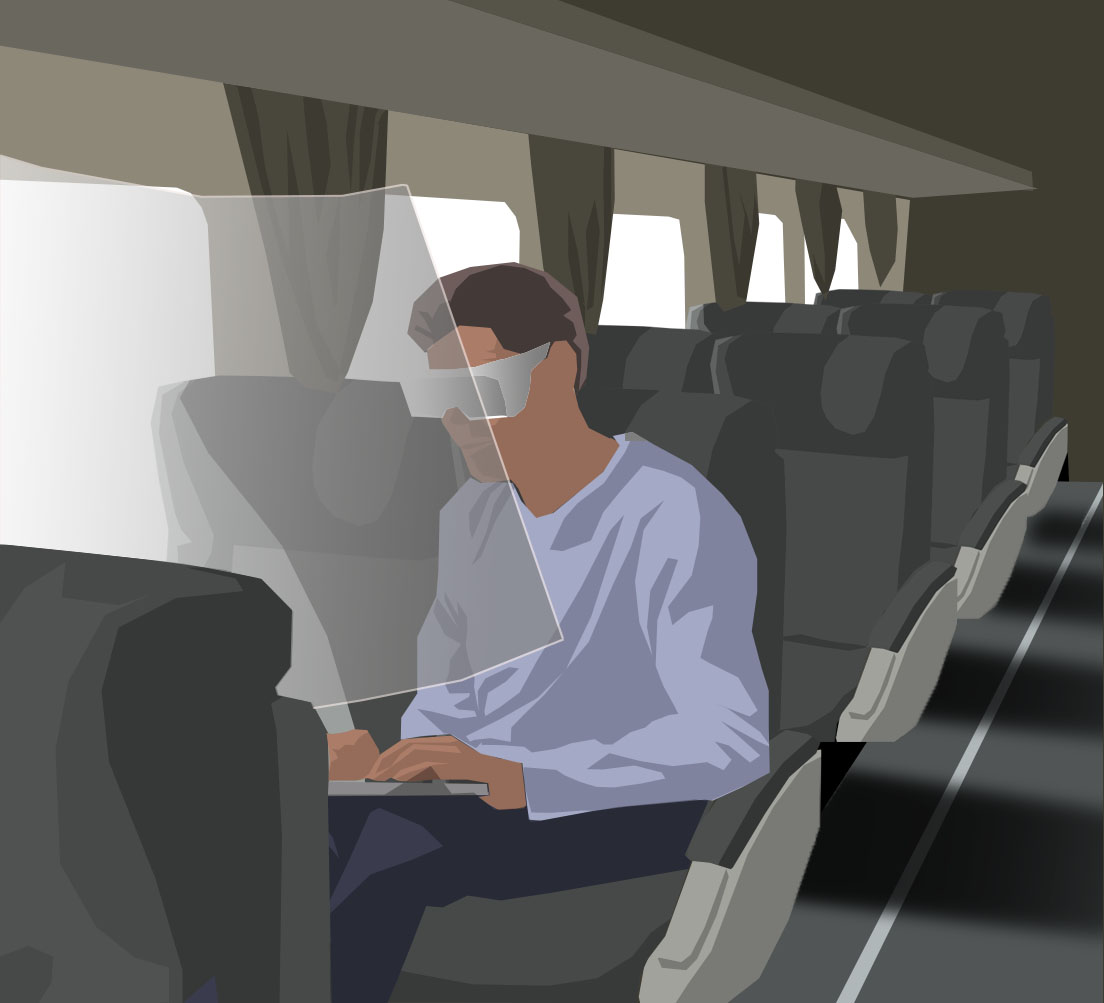}
	\caption{A vision of a portable virtual office enabled by VR.}
	\label{fig:vision}
\end{figure}

% POK: I think we need to rewrite the Introduction and keep it rather short. We can then go to Related Work as a separate section and weave in parts of the old introduction there. Before each Experiment we can then clearly explain why this experiment was carried out, is novel, and interesting.
Text entry in Virtual Reality (VR) is an important feature for many tasks, such as note taking, messaging and annotation. Existing consumer-grade VR systems, such as HTC Vive, Oculus Rift, Sony PSVR or Samsung's Gear VR, often rely on indirect control of a virtual pointer using hand-held controllers or head or gaze direction. However, these methods are limited in performance and consequently mostly used to enter short texts, such as passwords or names. Further, they require some degree of training due to their unfamiliarity to some users, compared to standard desktop and touchscreen keyboards that users are already familiar and proficient with. In addition, we believe a VR headset coupled with a keyboard can become an enabler for a full portable office in which a user can enjoy a motion-independent robust and immersive virtual office environment (Figure \ref{fig:vision}). For instance, users of touchdown spaces might find convenient to be able to carry their personalized large office configuration with multiple displays along with them in very tiny spaces.

However, while direct transplantation of standard keyboards to VR is viable, there are critical design parameters that should be investigated since it is plausible they affect performance.
In a companion paper \cite{grubert2018text} we investigate the performance of physical and touch keyboards and  physical/virtual co-location for VR text entry.

In this paper we focus specifically on investigating the method for virtually representing a user's hands in VR in several different ways. One possibility is to not reveal the hands at all, or resort to simply visually indicate to the user the actual pressed keys \cite{walker2017efficient}.

Alternatively, incorporating a video of the user's hand, within a VR world~\cite{mcgill2015dose} may be the closest representation to physical hands, at a possible cost of breaking the immersiveness of the virtual world. Another common representation that can be used to fit the VR style is to visualize the users' hands using a three-dimensional model, made to fit the scene style. However, depending on the level of sensing of the users finger motions, there may be visual differences between the look of the model and the real hands. Furthermore, the look of the rendered hand, a difference in gender, unnatural interpolation of motions may generate a dissonance between the user and the chosen avatar hands \cite{schwind2017these}. VR is not limited to physical limitations, and new representations may be proposed that better fit the text entry needs. For example, it is possible to visualize only the user's finger tips and thereby minimize visual clutter, leaving the keyboard mostly visible. %\JC{JG: SHOULD THE FOLLOWING SENTENCE GO SOMEWHERE ELSE?} We found this representation sufficient to transfer typing abilities to VR. Other possible representations may rendered hands totally opaque or semi-transparent to minimize occlusion.

\subsection{Contribution}

In this paper, we present the results of an experiment, in which we study the effect of four different hand representations: no hand representation, representing the hands via a three dimensional hand model, representing fingertips only, and representing the actual hands via blended video.%, closest to the view of the hands in reality. 

Our study indicates that while users tend to write comparably fast (with disabled correction options), the different representations lead to significant differences in error rate and preferences. Specifically, inverse kinematics hand model results in lower performance compared to a minimal fingertip visualization, while showing no hands at all results in comparable performance to a inverse kinematics hand model and  a minimal fingertip visualization results in comparable performance to a blended video view of the users physical hands.

\section{Related Work}

%\todo[inline]{discuss mcgill + brewster dose of reality, CHI 2016}
%\cite{mcgill2015dose}

There is a large body of work investigating interaction issues in VR ranging from perceptual issues \cite{Renner:2013:PED:2543581.2543590} to interaction tasks like navigation, spatial manipulation, system control or symbolic input \cite{Bowman:2004:UIT}.
%\hl{3DUI tasks introduction ending in symbolic input. 1-2 perceptual issues in VR, slater papers}
%\subsection{HCI studies in VR}
%\subsection{Symbolic Input in VR}

Text entry has been extensively researched (see~\cite{mackenzie2002text,zhai2004search,mackenzie_textentry,kristensson2015next,kristensson2009five} for surveys and overviews). Several strategies have developed in the text entry field to improve performance, in particular optimization \cite{mackenzie_soft,zhai2000metropolis,zhai2002movement,zhai2002performance,bi2010quasi,oulasvirta2013improving}, decoding (auto-correct) \cite{goodman_lm_soft,kristensson_relax,kristensson_five,weir_uncertain,vertanen_velocitap} and gesture keyboards \cite{zhai_shorthand,kristensson_shark,kristensson_thesis,zhai_word}.

\subsection{Text Entry in VR}
Relatively few text entry methods have been proposed for VR. Bowman et al.~\cite{bowman2002text} speculate that the reason for this was that symbolic input may seam inappropriate for the immersive VR, and a belief that speech will be the one natural technique for symbolic input. Indeed, when comparing available techniques, they found speech to be the fastest medium for text entry at about 14 words-per-minute (wpm), followed by using a tracked stylus to select characters on a tablet (up to 12 wpm), a specially dedicated glove that could sense a pinch gesture between the thumb and each finger at 6 wpm, and last a commercial chord keyboard which provided 4 wpm. While voice control is becoming a popular input modality \cite{pick2016swifter}, it has severe limitations of ambient noise sensitivity, privacy, and possible obtrusiveness in a shared environment \cite{Dobbelstein:2015,Tung:2015:UGI}. 
It has also been argued that speech may interfere with the cognitive processes of composing text \cite{shneiderman2000limits}. Furthermore, while dictation of text may feel natural, it is less so for text editing. Correcting speech recognition errors is also a challenge (see Vertanen~\cite{vertanen2009efficient} for a recent overview).

Prior work has investigated a variety of wearable gloves, where touching between the hand fingers may represent different characters or words, e.g., \cite{bowman2002text, Hsieh:2016:DWH, Kuester:2005, pratorius2015sensing}. Such devices enable a mobile, eyes-free text entry. However, most of them require a considerable learning effort, and may limit the user ability to use other input devices while interacting in VR, such as game controllers.%pratorius2014digitap

Yi et al.~\cite{Yi:2015} suggest a system that senses the motion of hands in the air, to simulate typing, claiming a rate of up to 29 wpm, measured not in VR, but while the users could see their hands, and keep them in front of the sensing device. Also, holding the hands in mid-air above some virtual plane is lacking any haptic or tactile sensation feedback and can become tiring quickly and may not fit long typing session.  

PalmType \cite{Wang:2015:PUP} uses a proximity sensor to use the non-preferred hand as a typing surface for the index finger of the preferred hand. The authors claim a rate which is better than touchscreen phones. However, the size of the simulated keyboard is limited to a hand size, and is hard to be rendered well in the current HMDs. It is also limiting the interaction to a single finger of a single hand, while the non-preferred hand is occupied as the type surface.

The mainstream mobile phone touchscreen keyboard might be a good text entry option \cite{gonzalez2009evaluation}. It is portable, and can generate a relatively high text entry rate \cite{reyal2015performance}. While we do not investigate this small type of touchscreen keyboard in this paper, we do believe it has potential. %\JC{\small \bf MP: It seems that mobile phones might be good for text entry when the user is moving a lot in the VR scene (alternative solution to strapping a slate on non-dominant arm), but for situations where the user is sitting and type text, a desktop keyboard or touch keyboard on table seems more efficient}.
Currently, a limitation with using a phone is its small size, which combined with the display resolution limitation does not generate a good experience. %Scaling up of the phone screen and the user's hands may elevate this problem, but this was outside the scope of this work.

\subsection{Keyboards for VR}
%\JC{{\bf EO:} Do we need to mention Walker et al. again with more details than the intro? }
Recent research has investigated the feasibility of typing on a physical full-sized keyboard (hereafter referred to as a desktop keyboard) in VR. An obvious problem is the lack of visual feedback. Without visual feedback users' typing performance degraded substantially. However, by blending video of the user's hands into virtual reality the adverse performance differential significantly reduced \cite{mcgill2015dose}.

Fundamentally there are three solution strategies for supporting keyboards in VR. First, by providing complete visual feedback by blending the user's hands into virtual reality. Second, by decoding (auto-correcting) the user's typing to compensate for noise induced by the lack of feedback. Third, by investigating hybrid approaches, such as minimal visual feedback, which may or may not require a decoder to compensate for any noise induced by the method.

Walker et al.~\cite{walker2016decoder} presented the results of a  study of typing on a desktop keyboard with the keyboard either visible or occluded, and while wearing a VR HMD with no keyboard display. They found that the character error rate (CER) was unacceptably high in the HMD condition (7.0\% average CER) but could be reduced to an average 3.5\% CER using an auto-correcting decoder. A year later, they showed that feedback of a  virtual keyboard in VR, showing committed types, can help users correct their hand positions and reduce error rates while typing \cite{walker2017efficient}.
They discovered that their participants typed at an average entry rates of 41.2--43.7 words per minute (wpm), with average character error rates of 8.3\%--11.8\%. These character error rates were reduced to approximately 2.6\%-4.0\% by auto-correcting the typing using the VelociTap decoder \cite{vertanen2015velocitap}.
In contrast, in this paper, we will show that by visualizing users' finger tips while typing, there is no need for an auto-correcting decoder as with the visual feedback users' character error rate is already sufficiently low for both desktop keyboard typing. This provides significant benefits to the user, as auto-correcting decoders, while useful for enabling quick and accurate typing, suffer from the auto-correct trap \cite{weir_uncertain}, which reduces entry rates and increases user frustration when the system inadvertently fails to identify the user's intended word.

McGill et al.~\cite{mcgill2015dose} investigated typing on a physical keyboard in Augmented Virtuality \cite{milgram1994taxonomy}. Specifically, they compared a full keyboard view in reality with a no keyboard condition, a partial and full blending condition. For the blending conditions the authors added a camera view of a partial or full scene into the virtual environment as a billboard without depth cues. They found, that providing a view of the keyboard (partial or full blending) has a positive effect on typing performance. Their implementation is restricted to typing with a monoscopic view of the keyboard and hands and the visualization of hand movements is bound by the update rate of the employed camera (typically 30 Hz). We study a complementary setup, in which we focus on virtual representations of the keyboard and hands. %While we focus on simple representations of fingers on purpose, future work could contrast the effects of different hand and keyboard visualizations (ranging from simple visualization like ours, over parametric hand models to blended hand and keyboard representations). 

Lin et al. \cite{lin2017visualizing} investigated the effects of different keyboard representations on user performance and preference for typing in VR but did not study different hand representations in depth.

Grubert et al. \cite{grubert2018text} investigated the performance of physical and touch keyboards and  physical/virtual co-location for VR text entry.

Schwind et al. \cite{schwind2017these} investigated the effects of gender on different virtual hand representations. The participants had to conduct various tasks, amongst them a typing task with a single sentence. The authors did not report any text entry metrics. For our experiment on hand representations, we reused their androgynous hand model as a compromise between male and female hand depictions.

\section{Hand Representation for Typing in VR}

In this work, we looked at the ability to enter a substantial amount of text in VR using a standard desktop keyboard setup, which is commonly available and requires little, if any, learning to use in VR. The existing keyboards already have comfortable form factors for two-hand interaction and provide the same haptic feedback in VR as they do in the real world. In contrast to typing in the real world, VR allows the user to be free of  physical limitations of this world. For example, if the user's hands occlude a keyboard, it is possible to make their virtual representation transparent so that the user can see the keyboard better. However, it is unclear if this would affect typing ability in VR.

%\JC{{\bf EO:} Any other new relevant papers in 2017? }

%\begin{figure}
%	\centering
%	\includegraphics[width=0.55\columnwidth]{images/dk2_disp.jpg}
%	\caption{Fresnel lens distortion and aliasing reduces the readability of text near the sides of the field of view}	\label{fig:dk2_disp}
%\end{figure}

%\todo[inline]{MP: Why useful: Maybe talk about the fact that we were interested in studying the hand representation because it could affect the sense of immersion of the experience (hands could fade into the VR experience) and the cognitive effort while typing (mainly effort to understand very quickly where each of the user's fingers are in relation to the keyboard at any time during typing). Btw, should we have some hypothesis here?}

%In this study, we wanted to compare the effects of common hand representation on the writing experience in VR using a standard keyboard. 
Various hand representations have already been proposed in prior work. However, it remains unclear how those could effect typing performance, effort and presence. Hence, we set out to compare common hand representations using a standard desktop keyboard. Specifically, we compare following hand representations: no hand representation as studied by Walker and Vertanen \cite{walker2017efficient}, blended video see-through of the user's hands as proposed by McGill et al. \cite{mcgill2015dose}, an inverse kinematic hand model as used by Schwind et al. \cite{schwind2017these} and a minimalistic sphere representation of the user's fingertips. Further, we conjecture that with appropriate visualization, there is no need for an auto-correcting decoder, which reduces the complexity of the typing system. 

\section{Experiment}
To investigate the effect of hand representation on typing in VR we carried out a controlled experiment. We used a within-subjects design with a single independent variable---\textsc{HandRepresentation}--with four levels: no hand representation (\textsc{NoHand}), inverse kinematic hand model (\textsc{IKHand}), fingertip visualization through spheres (\textsc{Fingertip}) and an Augmented Virtuality representation based on chroma keying (\textsc{VideoHand}), see Figure \ref{fig:conditionshandstudy}.

Since our objective was to evaluate the effect of hand representation on typing, our investigation primarily focused on statistically comparing typing performance metrics across the four conditions and generalizing these differences to the population. We therefore ensured we sampled participants with diverse study backgrounds and. importantly, we did not attempt to subsample participants with similar typing abilities.

Since text entry is a relatively complex task it is important to ensure participants are typing a sufficient amount of text for us to be able to accurately sample their true typing performance. Not doing so would introduce two unwanted sources of error: First, since the difficulty of typing individual phrases varies, inadequate typing inflates noise in the text entry and error rate measurements. Second, since participants are exposed to new unfamiliar conditions, there is inevitably a degree of learning and familiarization within the first few minutes within each condition.
For this reason we exposed every participant to 15 minutes of typing in each condition.

In addition, it is also important that the stimulus text models the text participants are likely to type. Such text is known as being \emph{in-domain}. For this reason we used stimulus text from a corpus derived from genuine emails typed on mobile devices \cite{vertanen2011versatile}.

\subsection{Method}
%\todo[inline]{MP: Figure 5 FINGERTIP shows yellow feedback for both hands. Is that intentional? I thought it would be blue and yellow?} 
In the condition \textsc{NoHand}, the participants saw no hand representation at all, but only the text they typed as well as green highlights of keys currently pressed. The condition \textsc{Fingertip}, showed semi-transparent yellow spheres at the finger tips but no other visualization in addition to the highlighted keys. The condition \textsc{VideoHand} used a blended billboard with a live video of the user's hands as well as the physical keyboard.  Please note, that in this condition the retroreflective markers where visible as well. While this might degrade the visual experience, the markers where kept as in the other conditions to avoid a potential confound. Also, there was an average end-to-end delay from finger movement to display in VR of 170 ms, which did not result in substantial coordination problems while typing as indicated by pre-tests.

The condition \textsc{IKHand} used an inverse kinematic hand model \cite{schwind2017these} as well as the key highlights.  

%The order of the conditions was counterbalanced. 
The experiment was carried out in a single 130-minute session structured as 5-minute welcoming and introduction, 5 minutes profiling phase, 15 minutes attachment of retroreflective markers, hand rigid bodies and calibration, a 90-minute testing phase (15 minutes per condition + ca. 7-minute breaks and questionnaires in between) and 15 minutes for final questionnaires, interviews and debriefing.

%In the \textsc{Reposition} condition, the keyboard and hands' would be spatially transformed such that they would be initially visible at the center of the user's field of view, and then fixed in space if users would move their heads. The four conditions are depicted in Figures \ref{fig:conditionsexternal} and \ref{fig:conditionsinternal}. The order of the conditions was counterbalanced. The experiment was carried out in a single 130-minute session structured as a five-minute introduction, 30 minutes of calibration data collection, an 80-minute testing phase (15 minutes per condition + five-minute breaks and questionnaires in between), and 15 minutes for final questionnaires, interviews and debriefing.

\subsection{Participants}
We recruited 25 participants from a university campus with diverse study backgrounds. %\todo[inline]{MP: Just to be clear, should we say here that the participants are all different than the ones for experiment 1?} 
All participants were familiar with QWERTZ desktop keyboard typing and QWERTZ touchscreen keyboard typing, none took part in the previous study. One participant had to be excluded due to logging issues. 
From the 24 remaining participants (14 female, 10 male, mean age 23.5 years, sd = 2.2, mean height 168.9 cm, sd = 36.7
), 15 indicated to have never used a VR HMD before, 4 to have worn a VR HMD once, 2 participants rarely but more than once and 3 participants to wear it occasionally. Seven participants indicated to not play video games, 2 once, 9 rarely, 2 occasionally, 1 frequently and 3 very frequently. Nineteen participants indicated to be highly efficient in typing on a physical keyboard, 3 to be medium efficient and 2 to write with low efficiency on a physical keyboard (we caution against over-interpreting these self-assessed performance indications). Six participants wore contact lenses or glasses. The volunteers have not participated in other VR typing experiments before.

\subsection{Apparatus and Materials}

\begin{figure}[]
	\centering
	\includegraphics[width=\columnwidth]{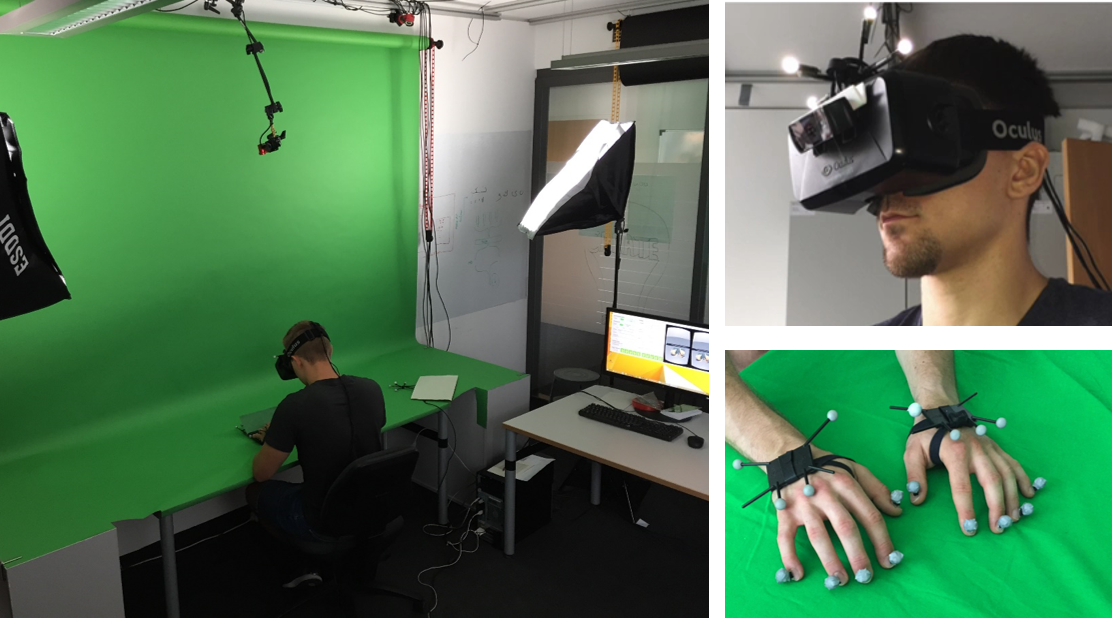}
	\caption{Left: Setup with green keying, tracking system and external lighting. Top right: HMD with external camera and tracking fiducial. Bottom right: fiducials used for tracking the hands in the  \textsc{IKHand} condition.}
	\label{fig:apparatushandstudy}
\end{figure}

Stimulus sentences were drawn from the mobile email phrase set \cite{vertanen2011versatile}, which provides a set of text entry stimulus sentences that have been verified to be both externally valid and easy to memorize for participants. Participants were shown stimulus phrases randomly drawn from the set.
%As shown in Figure \ref{fig:apparatushandstudy}, 
An OptiTrack Flex 13 outside-in tracking system was used for spatial tracking of finger tips and the HMD, again with a mean spatial accuracy of 0.2 mm. An Oculus Rift DK2 was used as HMD. A Logitech C910 camera (resolution 640x480, 30Hz) was mounted in front of the HMD and external lighting as well as a green keying background was installed to enable the \textsc{VideoHand} condition, see Figure \ref{fig:apparatushandstudy}. In addition to fiducials at the finger tips, another rigid-body fiducial was mounted at the back of the hands to support the \textsc{IKHand} condition. 

\begin{figure}[!b]
	\centering
	\includegraphics[width=1\columnwidth]{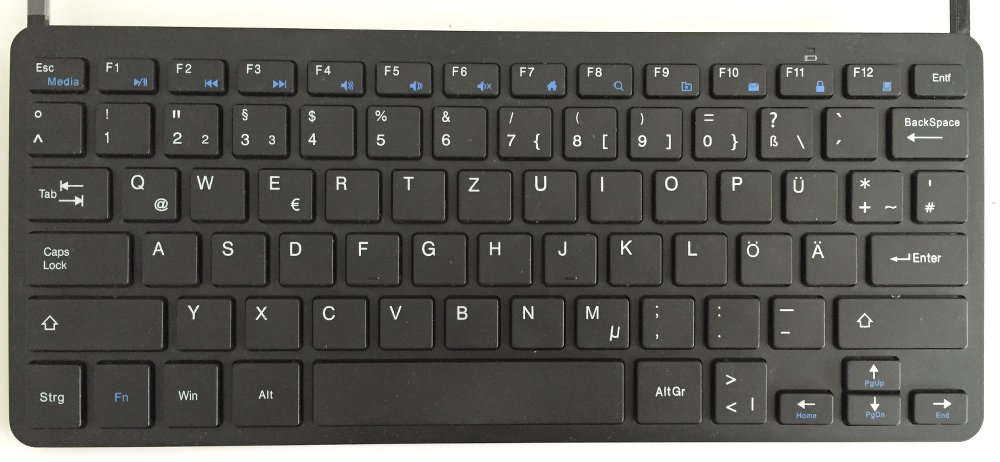}
	\caption{Physical keyboard used in the experiment.}
	\label{fig:keyboards}
\end{figure}

The physical keyboard was a CSL wireless keyboard %\todo[inline]{MP: Is CSL the brand? Is there a specific model type?} 
with physical dimensions of (width $\times$ height) ($w \times h$): 
272 $\times$ 92 mm and key dimensions of 15 $\times$ 14 mm, see Figure \ref{fig:keyboards}.

\subsection{Calibration Data Collection}

%\JC{LW: should we use cross references to our first study here? }
The calibration phase consisted of four parts: text entry profiling, interpupillary distance (IPD) calibration, finger tip calibration and finger length calibration. During the text entry profiling phase, participants were asked to copy prompted sentences using a desktop keyboard. Stimulus phrases were shown to the participants one at a time. Participants were asked to type them as quickly and as accurately as possible. Participants typed stimulus phrases for 5 min using a desktop keyboard. The IPD was determined with a ruler and then used for setting the correct camera distance for stereo rendering.

For finding out the IPD Participants were asked to sit upright and parallel to the experimenter holding a ruler below their eyes. With the experimenter having one eye closed, participants were asked to look straight into the open eye. The ruler was adjusted till its origin crossed the line between the open eye and the participants corresponding eye. The experimenter closed both eyes and opened the other one. Participants now looked again into the open eye while having read out their IPD by the experimenter.

This method is a fast and simple way to find out the IPD and was chosen to decrease the total time the participants need to wear the HMD. The IPD was then transferred to the experiment software.

For finger tracking, individual retroreflective markers were attached to the nails of participants using double-sided adhesive tape, see Figure \ref{fig:apparatushandstudy}, bottom right. The finger calibration aimed at determining the offset between the tracked 3D position of each finger tip and its corresponding nail-attached marker. To this end, the participants were asked to hit three soft buttons of decreasing size (large: 54 $\times$ 68 mm, medium: 35 $\times$ 50 mm, small: 15 $\times$ 15 mm) on a Nexus 10 touch surface. Initially, the virtual finger tips were shown at the registered 3D positions of the retroreflective markers. On touchdown, the virtual finger tips were transformed by the offset between the 3D coordinate of the touch point and the retroreflective marker. The final positions of the virtual finger tips were averaged across three measurements. Then the participants verified that they could actually hit targeted keys using their virtual finger tip. 
If necessary, the process was repeated. This calibration procedure was conducted for each finger individually. 

Additionally, rigid bodies holding four retroreflective markers were attached %pivotally
onto the back of each of the participant's hands. For each participant the hand's rigid body rotation was reset in the tracking system's software. 
Also, the information given by the hands' rigid bodies and the fingertips' markers was used to adapt the finger lengths of the inverse kinematic model to the participant's finger lengths. This ensured that the virtual fingertips were positioned near the real ones. 
Before each condition, the participants verified that they could actually hit targeted keys using their virtual finger tip. If necessary, parts of the calibration process were repeated.

\subsection{Procedure}
The order of the conditions was balanced across participants. In either condition, participants were shown a series of stimulus sentences. For an individual stimulus sentence, participants were asked to type it as quickly and as accurately as possible. Participants typed stimulus sentences for 15 minutes in each condition. The conditions were separated by a 5-minute break, in which participants filled out a the SSQ simulator questionnaire \cite{kennedy1993simulator}, the NASA TLX questionnaire \cite{hart1988development}, the IPQ \cite{regenbrecht2002real} spatial presence questionnaire and the Flow-Short-Scale \cite{rheinberg2003erfassung}.% as well as questions regarding their subjective assessment of the hand representations following the suggestions of Schwind et al. \cite{schwind2017these}.

Please note, that participants were not allowed to use the backspace key to correct errors. This was done in line with the suggestion by Walker and Vertanen \cite{walker2017efficient} to avoid excessive use of correction in the \textsc{NoHand} condition.

\subsection{Results}
Statistical significance tests for entry rate, error rate and time to first keypress (log-transformed) were carried out using General Linear Model (GLM) repeated measures analysis of variance (RM-ANOVA) with Holm-Bonferroni adjustments for multiple comparisons at an initial significance level $\alpha = 0.05$. Effect sizes for the GLM ($\eta_p^2$) are specified whenever they are available. All GLM analyses were checked for appropriateness against the dataset. We used GLM RM-ANOVA since it was appropriate for the dataset and provided more statistical power than  non-parametric tests.

Statistical significance tests for ratings and preferences were carried out using the non-parametric Friedman's tests coupled with Holm-Bonferroni adjusted post-hoc analyses with Wilcoxon signed-rank tests.

\subsubsection{Entry Rate and Time to First Keypress}
\begin{figure}
	\centering
	\includegraphics[width=\columnwidth]{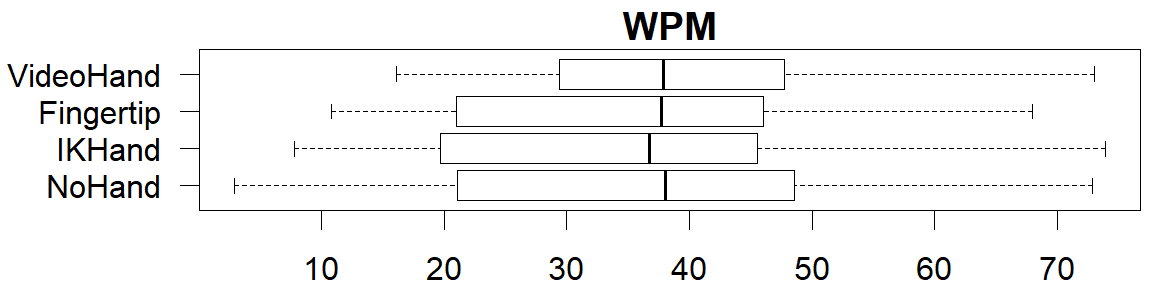}
	\includegraphics[width=\columnwidth]{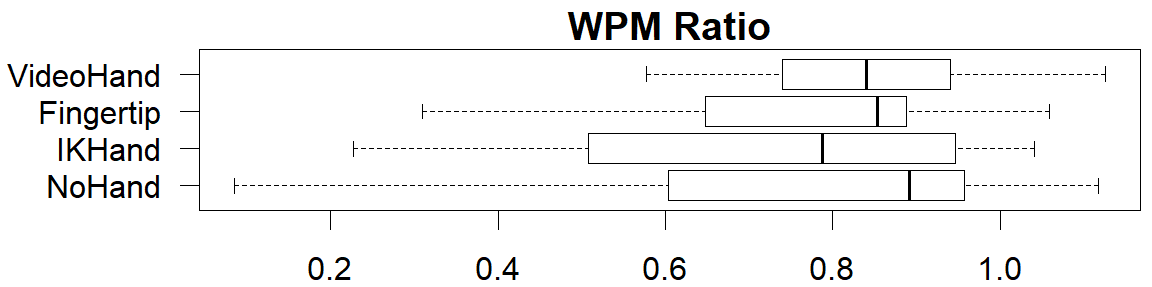}
	\includegraphics[width=\columnwidth]{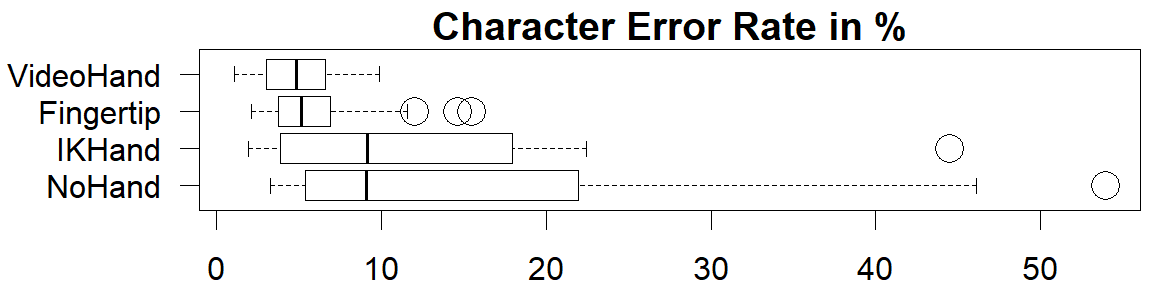}
	\includegraphics[width=\columnwidth]{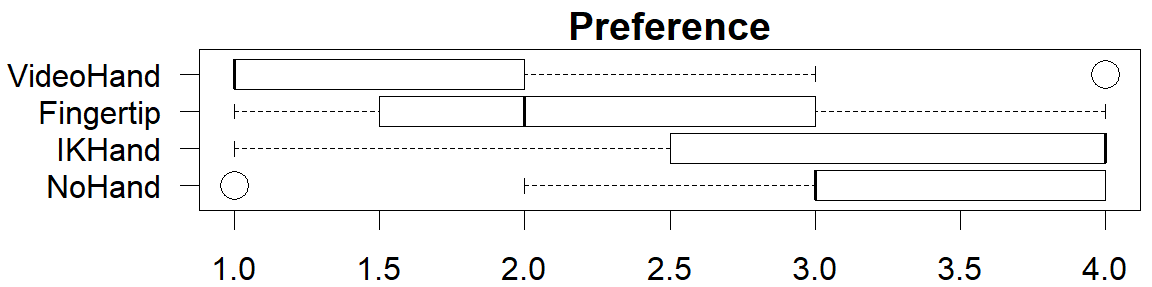}
	\caption{From top to bottom: Text entry rate in words per minute. Text entry ratio between condition and profiling phase. Character error rate.  Preference (1 = maximum preference best, 4 = minimum preference).}
	\label{fig:wpmhand}
\end{figure}

Entry rate was measured in wpm, with a word defined as five consecutive characters, including spaces; see Figure \ref{fig:wpmhand}, first row, for a graphical summary.
The entry rate was 36.1 wpm (sd = 18.1) for \textsc{NoHand}, 34.4 wpm (sd = 17.0) for \textsc{IKHand}, 36.4 wpm (sd = 15.3) for \textsc{Fingertip} and 38.7 wpm (sd = 13.6) for \textsc{VideoHand}, see Figure \ref{fig:wpmhand}, first row. The difference in entry rate was not significant ($F_{3,69} = 2.550$, $\eta^2_p = 0.1$, $p = 0.063$).

As a calibration point only, we also measured the entry rate ratio between the individual conditions and the profiling phase (Figure \ref{fig:wpmhand}, second row). On average, \textsc{NoHand} resulted in a 76\% entry rate compared to profiling, \textsc{IKHand} 72\%, \textsc{Fingertip} 78\% and \textsc{VideoHand} 84\%. Note that we intentionally did not control for typing proficiency when recruiting participants and it is therefore not meaningful to calculate statistical significance. We note that that the typing ratios are fairly high, indicating that the majority of the participants' typing abilities were preserved in VR. We conjecture this is due to participants being explicitly instructed not to perform corrections.

In addition we investigated the time to first keypress, a metric first suggested by McGill et al.~\cite{mcgill2015dose} to get an indication of the time it takes participants to orient themselves before typing the sentence.
The mean time to first keypress was 1.69 seconds (sd = 1.89 ) for \textsc{NoHand}, 1.57 seconds (sd = 1.42) for \textsc{IKHand}, 1.23 seconds (sd = 0.59) for \textsc{Fingertip} and 1.22 seconds (sd = 0.59) for \textsc{VideoHand}. A repeated measures analysis of variance on the log-transformed durations revealed that the differences were not statistically significant ($F_{3,69}=1.951$,$\eta^2_p=0.078$,$p=0.130$). 

Since the number of participants ($n =24$) was relatively high and the effect sizes are very low it is plausible there is no difference between the conditions for entry rate or time to first keypress.

%The difference between text entry rates between both experiments can be attributed to the experiment instructions. In the repositioning experiment users were able to correct phrases at will, in the hand representation experiment they were instructed not correct written phrases.

% \begin{figure}
% 	\centering
% 	 \includegraphics[width=0.8\columnwidth]{images/wpm-hand-24.png}
% 	 \caption{Text entry rate in words per minute for the hand representation experiment.}
% 	\label{fig:wpmhand}
% \end{figure}

% 
% \begin{figure}
% 	\centering
% 	 \includegraphics[width=0.8\columnwidth]{images/cer-hand-24.png}
% 	 \caption{Character error rate for the hand representation experiment.}
% 	\label{fig:cerhand}
% \end{figure}

\subsubsection{Error Rate}
Error rate was measured as character error rate (CER). %, word error rate (WER) and sentence error rate (SER). 
CER is the minimum number of character-level insertion, deletion and substitution operations required to transform the response text into the stimulus text, divided by the number of characters in the stimulus text. The character error rate (CER) was 15.2\% (sd = 14.2) for \textsc{NoHand}, 11.5\% (sd = 9.6) for \textsc{IKHand}, 6.3\% (sd = 3.7) for \textsc{Fingertip} and 5.1\% (sd = 2.5) for \textsc{VideoHand}, see Figure \ref{fig:wpmhand}, third row. An omnibus test revealed significance ($F_{3,69} = 9.029$, $\eta^2_p = 0.282$, $p < 0.001$). Holm-Bonferroni adjusted post-hoc testing revealed that there was no significant difference between \textsc{NoHand} and \textsc{IKHand} (adjusted p-value $< 0.025$). There was also no significant difference between \textsc{Fingertip} and \textsc{VideoHand}. However, there were significant differences between \textsc{NoHand} and both \textsc{Fingertip} and \textsc{VideoHand} and between \textsc{IKHand} and both \textsc{Fingertip} and \textsc{VideoHand} (adjusted $p < 0.025$). 

In other words, using no hand representation or an inverse kinematic hand model both resulted in a similar high CER. Using fingertip visualization with spheres or video see-through of the hand resulted in a similar relatively low CER. There is no statistically difference in CER between using fingertip visualization with spheres or video see-through of the hand.

\subsubsection{NASA-TLX, Simulator Sickness and Spatial Presence}
The overall median NASA-TLX rating was 57.9 for \textsc{NoHand}, 51.2 for \textsc{IKHand}, 47.5 for \textsc{Fingertip} and 43.8 for \textsc{VideoHand}. A Friedman's test revealed an overall significant difference ($\chi^2(3) = 10.399$, $p<0.05$). Holm-Bonferroni corrected post hoc analyses with Wilcoxon signed-rank tests revealed that the difference between \textsc{NoHand} and \textsc{VideoHand} was significant ($Z = -2.615$, $p < 0.01$). No other pairwise differences were significant.

%\subsubsection{Simulator Sickness}
The overall median nausea score rating was 2.0 for \textsc{NoHand} (oculo-motor: 7.0), 2.0 for \textsc{IKHand} (oculo-motor: 8.0), 1.0 for \textsc{Fingertip} (oculo-motor: 7.5) and 1.0 for \textsc{VideoHand} (oculo-motor: 6.5). There was no significant difference (Friedman's test; $\chi^2(3)=4.472$, $p=0.215$).

%\subsubsection{Spatial Presence}
%Figure \ref{fig:presencehand}, top, indicates the spatial presence ratings.
% \begin{figure}
% 	\centering
% 	% \includegraphics[width=\columnwidth]{images/presence-hand-24.png}
% 	 \includegraphics[width=\columnwidth]{images/pref-hand-24-2.png}
% 	% \caption{Hand representation experiment. Top: IPQ presence scores \cite{regenbrecht2002real} on a 7-item Likert scale (higher scores indicate a larger agreement with the item). O: Overall. S: Spatial Presence subscale. I: Involvement subscale. R: Experienced Realism subscale. Bottom: Preference ratings (1 is best, 4 is worst).}
% 	\caption{Hand representation experiment. Preference ratings (1 is best, 4 is worst).}
% 	\label{fig:presencehand}
% \end{figure}
For spatial presence, the median rating on a 7-item Likert scale was 3.5 for \textsc{NoHand}, 4.1 for \textsc{IKHand}, 3.8 for \textsc{Fingertip} and 4.0 for \textsc{VideoHand}. 
The differences for the overall rating were not statistically significant (Friedman's test; $\chi^2(3)=7.268$, $p=0.064$). However, using Friedman's test again, we found significant differences in the sub-scales Experienced Realism ($\chi^2(3)=8.442$, $p<0.05$) and Spatial Presence ($\chi^2(3)=9.846$, $p<0.05$). Involvement was not significant ($\chi^2(3)=0.872$,$p=0.872$). The only significant pairwise differences were between \textsc{NoHand} and all other conditions (Wilcoxon signed-rank tests).

\subsubsection{Preferences and Open Comments}
On a scale from 1 (best) to 4 (worst) the median preference rating for \textsc{NoHand} was 3, 4 for \textsc{IKHand}, 2 for \textsc{Fingertip} and 1 for \textsc{VideoHand}, see Figure \ref{fig:wpmhand}, bottom row. Friedman's test revealed an overall significant difference ($\chi^2(3)=30.150$, $p<0.0001$). Post-hoc analysis with Wilcoxon signed-rank tests and Holm-Bonferroni correction revealed there were no significance differences in preference between \textsc{NoHand} and \textsc{IKHand} and similarly no significant difference in preference between \textsc{Fingertip} and \textsc{VideoHand}. However, the difference in preference between \{\textsc{NoHand}, \textsc{IKHand}\} and \{\textsc{Fingertip}, \textsc{VideoHand}\} were significant ($p<0.005$).

Participants also commented about the reasons for their ratings. For the \textsc{NoHand} condition, two participants mentioned that it was hard or strenuous to orient themselves on the keyboard. However, three participants mentioned that showing no hand representation at all helped them to concentrate on writing. For \textsc{IKHand}, six participants mentioned that it was hard to orient on the keyboard, four mentioned that the hand was occluding too much of the keyboard, six participants experienced the hand as confusing or distracting. Two participants mentioned that the hands felt real, while two explicitly stated that they felt unreal with one stating that she would "prefer a more female version with nail polish". For \textsc{Fingertip}, six participants mentioned the feeling of accurate positioning, six that the spheres help to orient and do not occlude the keyboard substantially. One stated that the representation was "so abstract that it helped to orient better than the virtual hand model", one liked the "playful effect" of the spheres. However, one also mentioned that it felt "hard to orient if no finger is attached". For \textsc{VideoHand}, one participant highlighted the accurate depiction of hand positions, one that she felt in control, six mentioned that they prefer the representation as they are used to such a depiction of their hands, with one mentioning it would be the "least eerie choice". However, four participants also mentioned that the letters were hard to read due to the blurry depiction of the keyboard.

%\tfnusinodo[inline]{Jens continue}
%blind: hard to orient (1), helped to concentrate on writing (3), distracting (1), strenuous (1), worked better than expected (1)

%IK hands: confusing (3), felt real (2), hard to orient / position (6), distracting (3), felt unreal (1), ugly - would prefer a more female version with nail polish (1), occluding to much of the keyboard (4)

%spheres: accurate position (6), are not occluding and help to orient (6), were familiar (1), felt in control (2), helpful (1), hard to orient if no finger is attached (1), "were so abstract that they helped to orient better than the virtual hand model", playful effect (1)

%chroma: letters hard to read / blurry (4), accurate depiction of hand positions (1), felt in control (1), least strenuous (1), used to it / felt natural (6), least distracting (2), least eerie (1)

% \begin{figure}
% 	\centering
% 	 \includegraphics[width=\columnwidth]{images/pref-hand-24-2.png}
% 	 \caption{Preference ratings for the hand representation experiment (1 is best, 4 is worst).}
% 	\label{fig:prefhand}
% \end{figure}

%\subsection{Summary}

\section{Discussion}

Our research investigated the effects of different representations of the user's hands on text entry in VR. As a baseline, we used a video of the user's hands, composited in the virtual world as proposed by  McGill et al. \cite{mcgill2015dose}. In our experiment, which presented a sparse VR environment with minimalistic representations of a desk and a wall, it proved to be a useful representation. There could be cases where this representation could interfere with the virtual content rendering and may break immersiveness of the virtual experience.  
For example, in a futuristic settings, the user may be represented by an avatar wearing a spacesuit, while her hand will be represented by an everyday hands video. Even in a more "down-to-earth" setting, such as a virtual meeting or a virtual office, the video hand represents the conditions in the user's real environment and not in the VR set. Illumination, style, and for re-projection of the keyboard even the view direction of the video will not fit the VR world. 
Also, while using video hands does not require an external tracking system but merely a  RGB camera attached to the HMD (or integrated, as in HTC Vive), we as well as McGill et al. \cite{mcgill2015dose} needed to instrument the environment to enable robust chroma keying. For unprepared environments getting a robust image mask for both the user hands and the keyboard can be challenging (e.g. a dark office or home environment). While one could simply show the whole camera view, this potentially could reduce immersion further.

%Finally, the capturing of the user's hands video is dependent on the conditions in the real environment. It may be hard to capture a good video if, for example, the user is outdoor in the night, while it might be still possible to display avatar hands in the virtual world. 

Our study indicated that the studied minimalistic representation of finger tips has comparable performance to a video inlay. It has a high input rate with low error rate, and strong preference by the participants of our user study. We could imagine that this or other abstract representations could be beneficial over a video inlay for generic VR scenarios that aim at a high user presence. However, this should be investigated thoroughly in future work. %While, for our study, we used an external tracking system, HMD-based hand tracking solutions such as Leap Motion could potentially be used. However, their accuracy for typing tasks should be investigated first. 

A representation of the hands by a full 3D model, on the other hand, was found to lack in these areas. Its error rate was significantly higher, in fact, showing no significant differences to depicting no hands at all. One possible reason for the high error rate may be the low visibility of the keyboard behind the hands. Another may be due to any differences that may occur between the visible motion of the model and the actual motion of the user's hands. Since tracking of the user's hand is based many times on partial data (position is known only at recognizable markers on the finger in our case, or near recognizable features when computer vision is used), a fit of a model is used to interpolate the full motion and look of the hands. Specifically, while in our experiment the accuracy of finger tip positions in both \textsc{IKHand} and \textsc{Fingertip} conditions where the same, the other joints in the \textsc{IKHand} model where interpolated by the inverse kinematics model. Hence, they might show larger deviations from their physical counterparts. Any resulting difference in the model motion or look and the physical hands, may generate mistakes by the user, or even dissonance between the user and the avatar hands due to uncanny valley effects. As a result, our participants chose this representation as their least preferable. 

\subsection{Limitations and Future Work}

% \todo[inline]{MP: Should we also mention that a better representation to explore might include:
% - Display of the finger tips.
% - Semi-transparent display of the IK hands or VideoHand, just to hint the hands silhouette and identification of the fingers. The transparency may change along the hand, either total transparent in the palm and gradually becoming opaque to show the contour or transparent in the palm and gradual becoming opaque toward the finger tips.}

Our study focused on specific items in a large design space of how to design text entry systems in VR. For our evaluation, we focused on the  scenario of a user sitting in front of a desk doing extensive text entry. One reason was to measure text input rate at its limit. Another reason was the observation that this configuration is still popular by many VR applications that do not require the user to walk.  In particular, we can see a great potential of VR as a continuous unlimited VR display, replacing all physical screens in an office environment, supporting both 2D and 3D applications and visualizations. In this scenario, there is need for a robust text entry, which we believe can be filled by current keyboards with additional sensors for hand rendering.

% . Hence, We concentrated on a stationary typing experience at a desk studying peak performance of typing.
Alternatively, there are many mobile scenarios which could benefit from efficient text entry techniques, also for shorter text sequences. Here, either a handheld or arm-mounted touch screen might serve as a suitable interaction device. In this context, future work should investigate recent mobile text entry techniques for VR, e.g. based on gesturing  \cite{yeo2017investigating}. 

Also, we relied on high precision stationary optical tracking system. But even with this system, we did not sense the movement of physical key presses. The display of the fingers as as they move while typing may help people that do not touch type. The use of mobile depth sensors for hand tracking such as the leap motion could be a viable alternative, but their input accuracy for typing would need to be studied. %Besides analyzing head movements, future work could also investigate gaze patterns to study if less eye rotations occur in a repositioned keyboard visualization.

%Reducing latency and frame rate of sensing and display may increase typing speed.
%\item

Further, the four tested hand representations of our study are just four points in a vast possible design space, and we can imagine more to be suggested. For example, it may be that rendering a semi-transparent silhouette may combine the keyboard visibility of the finger tips with some hint of the hand model that may increase realism, but not too much to generate potential uncanny valley dissonance. In this regard, Logitech and Vive recently announced a developer kit for text entry using a physical keyboard, which employs a semi-transparent video inlay of the user's hands \cite{logitech17}. Also, the keyboards and hands were rendered at 1:1 scale in the VR world. However, this size limits the text visibility on the keys in the HMD display. Keyboards and fingers may be scaled up, allowing greater visibility, or even scaled down to create less occlusion. Future work should investigate the effects of this scaling on text entry performance as resolution and sharpness of VR HMDs keep increasing. Finally, we aim at studying the effects of abstract hand representations such as spheres compared to video inlays in more detail. Specifically, we want to measure presence of both representation in various VR scenarios that go beyond a neutral office environment (such as the previously mentioned space scenario). %\JC{\small \bf MP: Also has resolution of displays within HMD increases which makes text visibility better, we could also imagine scaling down the size of keyboard and hands for minimizing occlusion.} 

\section{Conclusions}

We have studied the effect of different representations of the user's hands on typing performance. We found that a minimalistic representation of the user's fingertips may enhance keyboard visibility and be as performant as viewing a live video of the user hands, while using 3D avatar hands that are fit to match the user hands, may decrese performance as much as not showing any view of the hands at all.

Finally, we believe that VR may expand from the current use of immersive experiences, to a work tool even in the common office, allowing information workers to interact and observe data without the limitation of physical screens. One barrier for such a vision is a robust text entry and editing tool, and we hope this work will be a step in this direction.

\section*{Acknowledgments}
Per Ola Kristensson was supported by EPSRC (grant number EP/N010558/1). We thank the volunteers in the experiment and the anonymous reviewers for their feedback.

\balance{}

\bibliographystyle{abbrv}
%%use following if all content of bibtex file should be shown
%\nocite{*}
\bibliography{hands}
\end{document}